\documentclass[fleqn, usenatbib]{mnras}
\usepackage[T1]{fontenc}
\usepackage{ae,aecompl}
\usepackage{graphicx}
\usepackage{color}
\usepackage{amsmath}
\usepackage{tablefootnote}
\usepackage{multirow}
\usepackage{threeparttable}

\title[Schwarzschild model of Fornax dSph]{Schwarzschild dynamical model of the Fornax dwarf spheroidal galaxy}

\author[K. Kowalczyk et al.]{
Klaudia~Kowalczyk$^{1}$\thanks{E-mail: \href{mailto:klaudia.kowalczyk@gmail.com}{klaudia.kowalczyk@gmail.com}},
Andr\'es del Pino$^{2}$,
Ewa L.~{\L}okas$^{1}$
and Monica~Valluri$^{3}$
\\
$^{1}$Nicolaus Copernicus Astronomical Center, Polish Academy of Sciences, Bartycka 18, 00--716 Warsaw, Poland\\
$^{2}$Space Telescope Science Institute, 3700 San Martin Drive, Baltimore, MD 21218, USA\\
$^{3}$Department of Astronomy, University of Michigan, 1085 South University Ave., Ann Arbor, MI 48109, USA
}

\pubyear{2018}

\begin{document}
\label{firstpage}
\pagerange{\pageref{firstpage}--\pageref{lastpage}}
\maketitle

\begin{abstract}
We present a full dynamical model of the Fornax dwarf spheroidal galaxy obtained with the spherically symmetric
Schwarzschild orbit superposition method applied to the largest kinematic data set presently available. We
modelled the total mass content of the dwarf with the mass-to-light ratio $\Upsilon$ varying with radius and found that
Fornax is with high probability embedded in a massive and extended dark matter halo. We estimated the total mass
contained within 1\,kpc to be $M(<1\,$kpc$)=1.25^{+0.06}_{-0.13} \times 10^8\,$M$_{\sun}$.
The data are consistent with
the constant mass-to-light ratio, i.e. the mass-follows-light model, only at 3$\sigma$ level, but still require a high
value of $\Upsilon \approx 11.2\,$M$_{\sun}/$L$_{\sun}$. Our results are in general agreement with previous estimates
of the dynamical mass profile of Fornax. As the Schwarzschild method does not require any assumptions on the orbital
anisotropy of the stars, we obtained a profile of the anisotropy parameter $\beta$ as an output of our modelling. The
derived anisotropy is close to zero in the centre of the galaxy and decreases with radius, but remains
consistent with isotropic orbits at all radii at 1$\sigma$ confidence level.
\end{abstract}


\section{Introduction}
\label{intro}

The distribution of mass in galaxies remains a subject of lively debate between astrophysicists supporting the
existence of dark matter and those favouring various implementations of modified gravity.
The most significant discrepancies between the mass profiles inferred from kinematics of the tracer and the distribution
of light, which in this paper we will interpret as indications of high dark matter content, are found in
galaxies belonging to the class referred to as dwarf spheroidals (dSph). Thanks to their proximity, dSphs of the Local
Group (LG) are the focus of attention in studies of dark matter distribution. Resolving single stars, both in
photometric and spectroscopic observations, gives astronomers a unique opportunity to model these galaxies in detail,
which at the moment seems to be limited mainly by the relatively small numbers of measurements.

The most numerous data samples, both photometric and spectroscopic, are currently available for the Fornax dSph, the
second largest and most luminous (after the Sagittarius dSph) satellite of the Milky Way. As a result, over the last
two decades the dynamics of Fornax has been modelled by many authors using various methods.
The simplest approach
takes advantage of estimators of dynamical mass roughly independent of the velocity anisotropy of the tracer, measured
at the half-light radius (or another predefined scale; \citealt{walker_2007}; \citealt{walker_2009b};
\citealt{walker_2011}; \citealt{amorisco_2013}).
Interestingly, \citet{walker_2011} and \citet{amorisco_2013} identified multiple stellar populations of different
metallicities as well as spatial distribution and kinematics which allowed them to measure the mass at various radii.

More sophisticated methods aim to reproduce the full mass profile of the galaxy. The most widely used modelling based on
solving Jeans equations \citep{GD} suffers from the mass-anisotropy degeneracy \citep{binney_1982} and therefore
requires additional assumptions concerning the anisotropy (e.g. constant with radius or of particular functional form)
and/or mass profiles (e.g. mass follows light, a particular density profile of the dark matter halo).
The Fornax dwarf has been modelled with various combinations of assumptions by \citet{lokas_2002}, \citet{wang_2005},
\citet{klimentowski_2007}, \citet{strigari_2008}, \citet{lokas_2009}, \citet{hayashi_2012}, \citet{hayashi_2015},  \citet{diakogiannis_2017}
and \citet{read_2018}. We can also find in the literature phase-space \citep{amorisco_2011} and action-based
\citep{pascale_2018} models as well as those applying the Schwarzschild \citep{schwarzschild_1979} orbit superposition
method \citep{breddels_2013b, jardel_2012, jardel_2013}.

To give just an order of magnitude, the dynamical mass within the half-light radius of 0.71\,kpc (at the adopted
distance of 147\,kpc) is estimated to be $5.6\times 10^7$M$_{\sun}$ \citep{mcconnachie_2012}, with the mass-to-light
ratio $\sim 10 $M$_{\sun}$/L$_{\sun}$, which suggests that Fornax is embedded in a massive dark matter halo. The
galaxy is also elongated: the projected ellipticity $\epsilon=1-b_p/a_p=0.30\pm 0.01$, where $b_p$ and $a_p$ are the
projected semi-minor and semi-major axis, respectively \citep{irwin_1995}. It shows traces of a major merger about 6
Gyr ago \citep{delpino_2015} and probably did not interact strongly with the Milky Way due to its extended orbit
\citep{battaglia_2015} so has no strong tidal features and can be assumed to be in dynamical equilibrium.
Additionally, the galaxy contains little to no gas \citep{bouchard_2006}, which may appear surprising in the light of
the above since a most likely mechanism to strip the gas, namely the ram pressure, would require at least one close
approach to the Milky Way.

Although the observations of LG dwarfs evince their non-negligible ellipticities \citep{mateo_1998,
mcconnachie_2012}, most studies still treat them as spherically symmetric. We can find in the literature only a few
attempts of generalizing shapes of galaxies for various purposes. These studies include: full dynamical modelling
(\citealt{jardel_2012, jardel_2013, jardel_2013a} assuming axisymmetric edge-on stellar component and spherical dark
matter halo; \citealt{hayashi_2012, hayashi_2015} assuming both components to be axisymmetric), calculations of
corrections to possible dark matter annihilation fluxes (\citealt{sanders_2016} obtaining values of mass with
strong assumption or from estimators) or tracing the evolution of a satellite (\citealt{sanders_2018} using predefined
density profiles). Unfortunately, up to date there has been no research done on mock objects showing that the intrinsic
shapes and therefore multiple free parameters (up to 6 for \citealt{hayashi_2015}) can be reliably derived with
currently available data samples.

In this paper we present a new full dynamical model of the Fornax dwarf obtained using the spherically symmetric
Schwarzschild orbit
superposition method in the form developed in \citet{kowalczyk_2017, kowalczyk_2018}. In \citet{kowalczyk_2018} we
used mock data selected from the outcome of an $N$-body simulation of a major merger of two disky dwarf galaxies that
led to the formation of a dark matter dominated, spheroidal galaxy similar to Fornax. We examined the effect of its
non-spherical shape on the inferred properties and quantified the systematic errors inherent in the method. The
results of this previous study and the similarities between the mock object and the real Fornax will help us interpret
the outcome of the present work.

The paper is organized as follows. In Section~\ref{data} we present the properties of Fornax and our observational data
sets whereas in Section~\ref{model} and \ref{results} we describe the modelling procedures and the obtained results. In
Section~\ref{discussion} we summarize our findings and extensively compare them to those available in the literature.

\section{Data}
\label{data}

In this section we give a general overview of our observational data samples, both photometric and spectroscopic.
The detailed description of the measurements and procedures used in merging various catalogues can be found in
\citet{delpino_2015} and \citet{delpino_2017}. In Table~\ref{tab:fornax_param} we present the observational parameters
of Fornax which we assume in this paper: the coordinates of the centre, the distance and relative velocity with respect
to the observer and the total luminosity. For consistency with the data sets which will be described in the next
subsections we use the values given in \citet{delpino_2013} and references therein.

\begin{table}
\begin{center}
 \caption{Observational parameters of the Fornax dSph (\citealt{delpino_2013} and references therein).}
\label{tab:fornax_param}\ \\
\begin{tabular}{lc}
 \hline
quantity&\ \ \ \ value\ \ \ \ \\
\hline
RA $\alpha$ (J2000.0)&2h$\,39\arcmin\,53.1\arcsec$\\
Dec $\delta$ (J2000.0)&$-34^{\circ} \,30\arcmin\,16.0\arcsec$\\
heliocentric distance [kpc]&136\,$\pm$\,5\\
heliocentric velocity [km\,\,s$^{-1}$]&55.3\,$\pm$\,0.1\\
luminosity $L_V\,[$L$_{\sun}]$&$1.55\times 10^7$\\
\hline
\end{tabular}
\end{center}
\end{table}

\subsection{Photometry}
\label{data_photo}

Our photometric sample is an extensive ensemble of archival data with more than $3.5\times10^{5}$ stars reaching
V$\sim$23.5 with the completeness of 50\% based on the calibrated photometry obtained by \citet{stetson_2000},
\citet{stetson_2005} and \citet{deboer_2012}. It covers the main body of the galaxy extending up to more than $\sim
1.2^\circ$ to the north-east from the centre. In order to ensure that all stellar magnitudes are in the
same photometric system, a very precise internal calibration between the catalogues was applied. This calibration
was performed using the best measured stars common between the catalogues (18.5 < V < 23.5), by an iterative
second-order fitting between their sky coordinates, magnitudes and colour. After the calibration, the catalogues were
merged, keeping all the non-common stars between the catalogues as well as only the best-measured magnitudes for the
common ones. Finally, we cleaned our photometry from probable non-stellar objects and stars with high magnitude errors
by using quality flags provided by {\sc Daophot} and {\sc Dophot}. A detailed description of the whole procedure
together with the assumed acceptable ranges for the quality flags can be found in \citet{delpino_2015}.

As shown by \citet{delpino_2015}, the spatial distributions of stars belonging to different stellar populations differ
significantly. Therefore, we decided to limit the sample used for the derivation of the luminosity profile to stars of
the red giant branch (RGB), the horizontal branch (HB) and the red clump (RC) as the spectroscopy is usually done for
these types of objects. In order to guarantee high (90\%) completeness of our sample we considered only stars with
V<22, additionally excluding the stars: of the main sequence, strongly reddened (on the right-hand side of the RGB) and
brighter than RGB and HB. The selected sample contained 65\,797 stars and for simplicity we will refer to them as RGB.

\begin{figure}
\begin{center}
\includegraphics[width=\columnwidth, trim=0 0 0 0, clip]{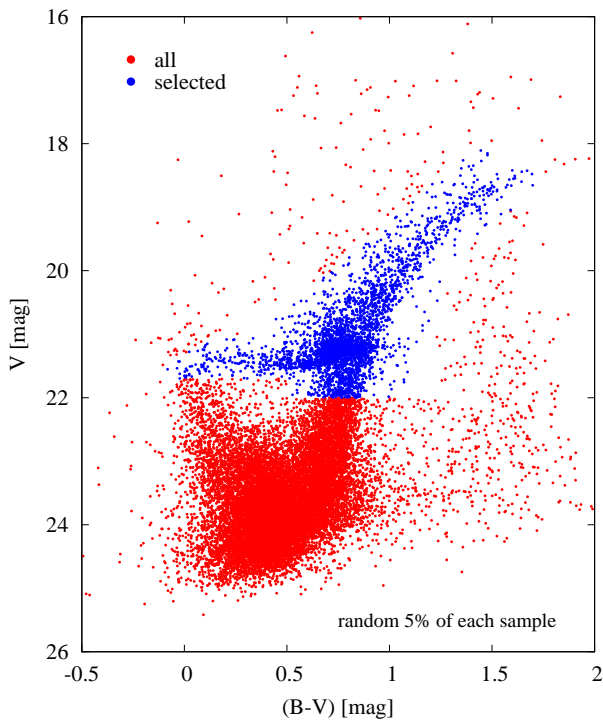}
\caption{Colour-magnitude diagram for the stars in the Fornax dwarf
spheroidal. In red we present random 5\% of all available stars with photometric measurements and in blue 5\% of the
stars selected for the luminosity profile (RGB, HB, RC).}
\label{fig:cmd}
\end{center}
\end{figure}

We present the comparison of the full sample (in red) and the stars chosen for the measurement of the luminosity
profile (in blue) in the colour-magnitude diagram of Figure~\ref{fig:cmd}. To avoid overcrowding only random 5\%
of each sample is shown.

\begin{figure}
\begin{center}
\includegraphics[width=0.8\columnwidth, trim=0 0 175 0, clip]{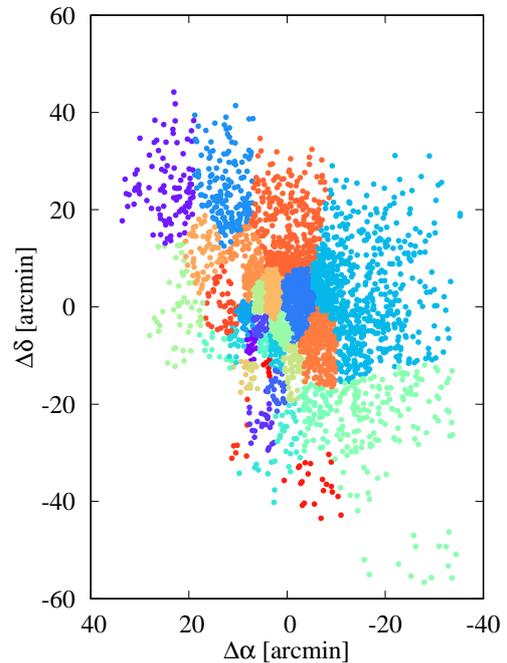}
\caption{The last iteration of the Voronoi tessellation for the stars with
line-of-sight velocities. Each dot represents a star and colours denote different Voronoi cells.}
\label{fig:cells}
\end{center}
\end{figure}

\begin{figure}
\begin{center}
\includegraphics[width=\columnwidth, trim=0 0 0 0, clip]{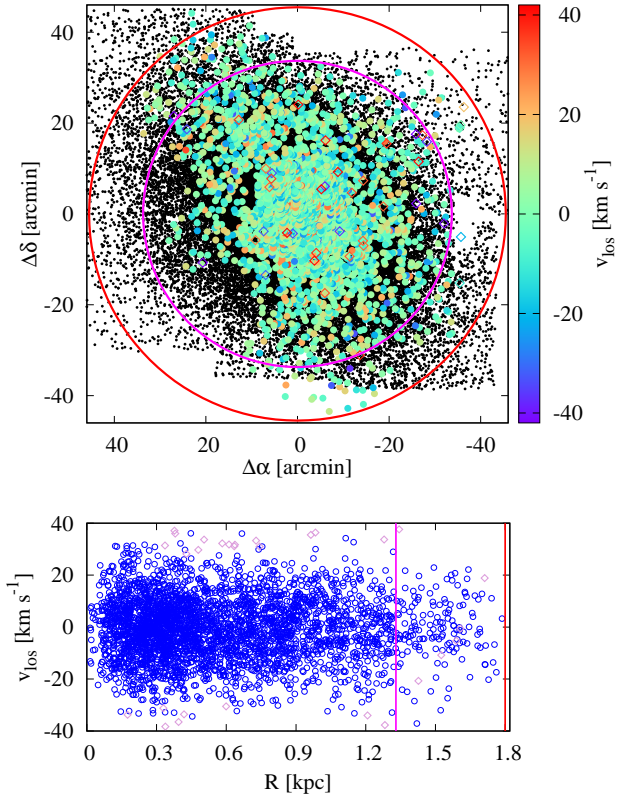}
\caption{\textit{Top panel:} Two-dimensional map of the Fornax dSph. Small black
dots represent the sample of stars chosen for the luminosity profile. With large dots we overplot the kinematic data
colour-coded with the line-of-sight velocity corrected for the bulk movement of the galaxy as a whole. Diamonds show
stars rejected as a result of the Voronoi tessellation. With coloured circles (and vertical lines in the bottom panel)
we present the outer radii chosen for the data sets used in modelling: luminosity profile (magenta) and moments of the
line-of-sight velocity (red).
\textit{Bottom panel:} Line-of-sight velocity measurements as a function of
distance from the centre of the galaxy. Blue circles represent stars used in the modelling whereas pink diamonds show
stars rejected by the Voronoi tessellation.}
\label{fig:fornax}
\end{center}
\end{figure}

\subsection{Spectroscopy}
\label{data_spec}

The spectroscopic list of stars was obtained after combining the catalogues with the largest number of spectroscopic
measurements for RGB stars in Fornax \citep{battaglia_2006, walker_2009a, kirby_2010, letarte_2010}. The procedure
followed to combine all catalogues is explained in detail in \citet{delpino_2017}. For the present work only good
precision in the line-of-sight velocity of the stars was required which allowed us to include some of the Fornax stars
with poorly determined chemistry but reliable kinematic measurements. Some of the stars were common to two or more
catalogues and in such cases we adopted the weighted mean of all the measurements (with weights determined by the
errors) for the line-of-sight velocity of the star.

The resulting spectroscopic list was cleaned from outliers and doubtful member stars with the Voronoi tessellation
technique. Using this method, an adaptive grid is set over the stellar coordinates, adjusting the sizes of the cells to
keep a constant number of stars within each cell. Consequently, areas with lower number of observed stars are covered
with larger grid cells. The tessellation was performed in an iterative way until convergence, keeping $\sim$80 stars
per cell in each iteration and removing from every cell stars lying outside the 3$\sigma$ of the line-of-sight
velocity distribution within the cell. In Figure~\ref{fig:cells} we present the last iteration of the Voronoi
cleaning. Dots represent the total of 3\,286 stars kept as highly probable members of Fornax and are colour-coded
according to the cell to which they were assigned.

We note that some of the cells end up with less stars than our target number of 80 stars. This is due to the lack of
sampled stars in the boundaries of the cells that causes the process of tessellation to stop before reaching the
desired number of stars. This does not cause any problem when considering individual cells since 10 or more
stars is sufficient for screening out clear contaminants.

\subsection{Final data set}

In Figure~\ref{fig:fornax} we present the final data set which we will use for the modelling in Section~\ref{model}.
The top panel shows a two dimensional map of the sky centred on the coordinates given in Table~\ref{tab:fornax_param}.
Small black dots denote all RGB stars described in Section~\ref{data_photo} whereas large dots represent the stars
with spectroscopic measurements, colour-coded with the line-of-sight velocity relative to the mean velocity of Fornax
(Section~\ref{data_spec}). Circles indicate the outer boundaries of the luminosity profile (magenta) and kinematic
data (red). We will justify the selection of those boundaries in Section~\ref{model}.

The kinematic data (line-of-sight velocities) as a function of the physical distance from the centre of Fornax are
presented in the bottom panel of Figure~\ref{fig:fornax} with blue circles. In both panels we show the stars rejected
during the Voronoi cleaning with diamonds.

\section{Modelling}
\label{model}

\begin{figure}
	\begin{center}
\includegraphics[width=0.9\columnwidth, trim=0 0 0 0, clip]{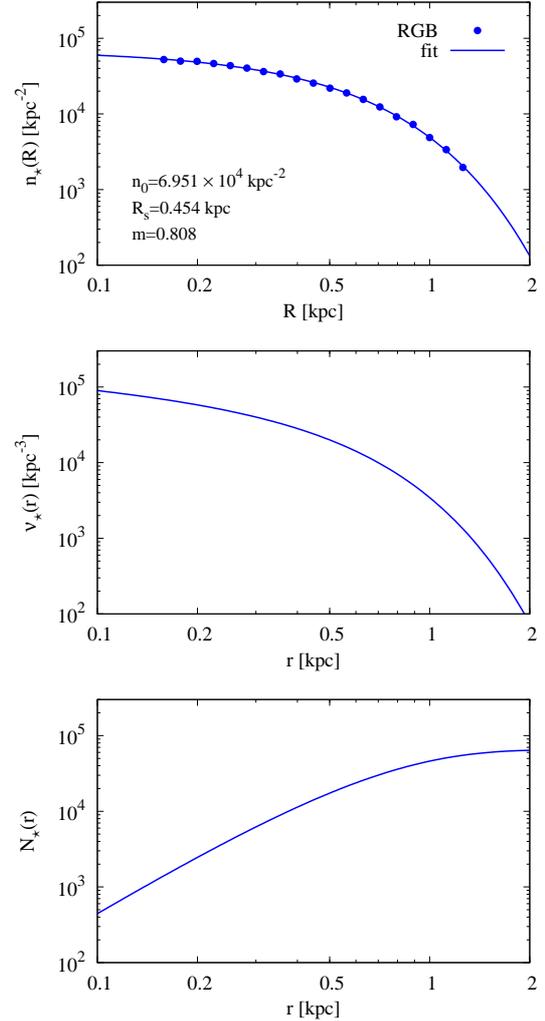}
\caption{\textit{Top panel:} The luminosity profile of the Fornax dSph. Blue
dots represent the measurements based on the RGB stars whereas the line shows the best-fitting S\'ersic profile.
The parameters of the fit are quoted in the bottom left corner.
\textit{Middle and bottom panels:} The 3D deprojected density and cumulative stellar number
profiles assuming the parameters of the best-fitting S\'ersic distribution.}
\label{fig:niu_M}
\end{center}
\end{figure}

In this section we describe the modelling procedures applied in order to derive the density and anisotropy profiles of
the Fornax dSph. In the top panel of Figure~\ref{fig:niu_M} we present the luminosity profile of Fornax
in terms of stellar density where points
indicate the measurements from the data described in Section~\ref{data_photo} while the solid line shows the
best-fitting S\'ersic profile \citep{sersic_1968} given as:
\begin{equation}
	\label{eq:sersic_2D}
	n_{\star}(R)=n_0\,{\rm exp}[-(R/R_{\rm s})^{1/m}],
\end{equation}
where $n_0$ is the normalization, $R_{\rm s}$ is the characteristic radius and $m$ is the S\'ersic index.
The parameters of the best-fitting profile are:
$n_0=6.951\times 10^4$\,kpc$^{-2}$, $R_{\rm s}=0.454$\,kpc, $m=0.808$ and they are also cited in the figure.
As the uncertainties of our dynamical modelling (see Sec.\,\ref{results}) will be driven mostly by
the sampling errors of kinematics, we will assume that the fit is very precise and will not include errors associated
with it in further analysis.

The inner radius of the profile, i.e. the radial distance of the innermost data point, was set to $r =0.16$\,kpc in
order to avoid overcrowding, i.e. the artificial underestimation of values due to the insufficient spatial resolution
of an instrument, while for the outer radius we adopted $r =1.26$\,kpc, the maximum value of the radius of the
circle with full coverage by data fields (see Figure~\ref{fig:fornax}).

We obtained the 3D stellar density $\nu_{\star}(r)$ and the cumulative stellar number profiles $N_{\star}(r)$ by
deprojecting the best-fitting S\'ersic
profile with the analytical formulae \citep{lima_1999} and we present them in the middle and bottom panels
of Figure~\ref{fig:niu_M}, respectively. The total number
of stars resulting from the fit is $N_{\rm s}$=65\,235.3.

We modelled the data by applying the spherically symmetric Schwarzschild orbit superposition method
which we have adopted for dwarf spheroidals of the Local Group and tested on mock data in
\citet{kowalczyk_2017} and \citet{kowalczyk_2018}. As in previous studies we used 10 bins linearly
spaced in radius and set the outer radius to $r =1.8$\,kpc. Such a value was a compromise between using all
velocity measurements not rejected by the cleaning procedure (see Section~\ref{data_spec}) and having a reasonable
number of stars in the outermost bin in order to maintain satisfying statistics.

We quantify the projected distribution of stellar mass in Fornax as a fraction of mass in a given bin $M_l(R)$
under the assumption of equal masses of stars.
The fraction of stellar mass was calculated as a ratio between the number of stars in a bin and the total number
derived from the parameters of the best-fitting S\'ersic profile. Since the luminosity profile is extrapolated for
small ($r < 0.16$\,kpc) and large ($r> 1.26$\,kpc) radii, fractions in the innermost and two outermost bins were
obtained `theoretically', i.e. estimating the number of stars by integrating the fitted profile. Moreover, the
kinematics of the data set was expressed in terms of proper moments of the line-of-sight velocity: the second
($m_2$), third ($m_3$) and fourth ($m_4$), calculated with estimators based on the sample of $N$ line-of-sight
velocity measurements $v_i$ \citep{lokas_2003}.

\begin{figure}
\begin{center}
\includegraphics[width=\columnwidth, trim=0 0 0 0, clip]{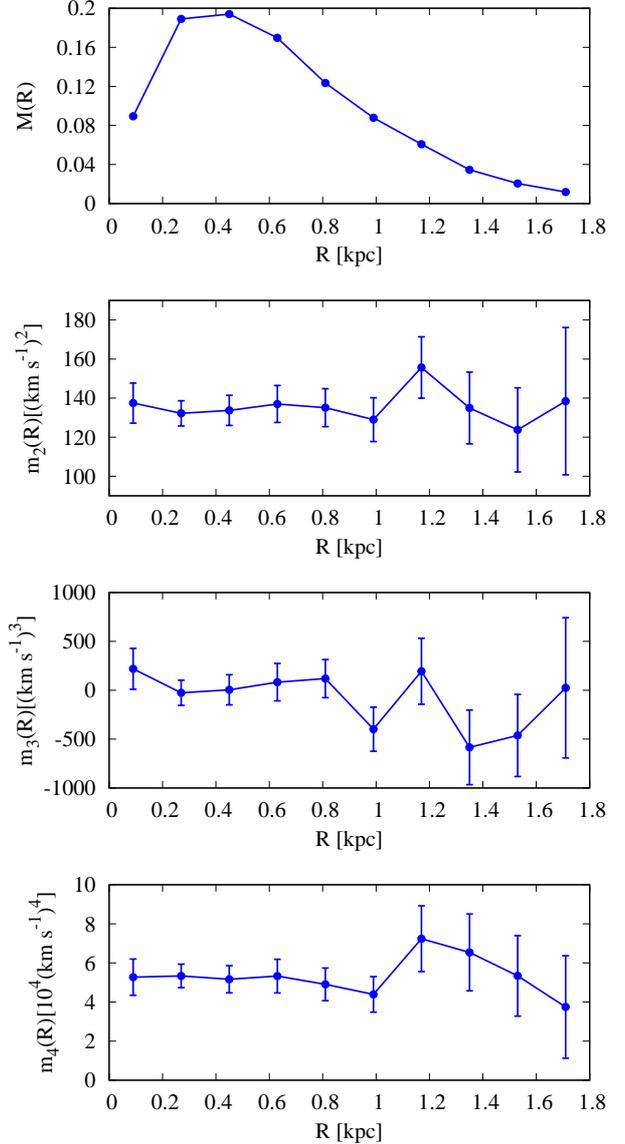}
\caption[Mass fraction and velocity moments profiles of the Fornax dSph]{Projected mass fraction and the
2nd, 3rd and 4th velocity moment (in consecutive panels) of the Fornax dSph calculated in 10 radial bins.
Error bars represent 1$\sigma$ sampling errors.}
\label{fig:obs_fornax}
\end{center}
\end{figure}

The profiles of the fraction of mass and line-of-sight velocity moments are presented in the consecutive panels
of Figure~\ref{fig:obs_fornax} where the error bars denote the sampling
errors. For the fraction of mass we adopted Poissonian errors (which in Figure~\ref{fig:obs_fornax} are smaller than
the data points) whereas the errors for the line-of-sight velocity moments were calculated analytically assuming normal
parent distributions \citep{kendall_1977}.

We modelled the total mass distribution in Fornax with the local mass-to-light ratio:
\begin{equation}
\label{eq:local_upsilon}
 \Upsilon (r)=\frac{\nu_{\rm tot}(r)}{{\nu}_{\star}(r)},
\end{equation}
varying with radius from the centre of the galaxy following a cubic function in log-log scale:
\begin{equation}
\label{eq:Upsilon}
\log\Upsilon(r) = \left\{
\begin{array}{ll}
\log \Upsilon_0 & \rm{for}\,\,\, r< r_0\\
a(\log r - \log r_0)^3 +\log \Upsilon_0 & \rm{for}\,\,\, r> r_0\\
\end{array} \right.
\end{equation}
where $\Upsilon(r)$ is dimensionless and $r$ is given in kpc.
Such a function has been first introduced
by \citet{kowalczyk_2018} as a satisfactory approximation of the true profile derived for their
mock Fornax dSph analogue when limiting the 
fitting to two free parameters, $a$ and $\Upsilon_0$,
which are constants defining the density model. We assume that in the centre the mass-to-light ratio is
constant, i.e. that mass follows light as the derivation
of the central slope is rather impossible with the current data. We set the minimum of the cubic curve to $\log r_0=-0.8$ as
a consequence of the adopted value of the innermost datapoint for the luminosity profile.

In order to express $\Upsilon_0$ in the convenient units of the solar mass-to-light ratio we assumed that the whole
stellar mass in Fornax is distributed in the same way as our RGB sample and translated the total luminosity
(Table~\ref{tab:fornax_param}) to the total stellar mass with $\Upsilon_{\star} = 1$\,M$_{\sun}/$L$_{\sun}$. Therefore,
$\Upsilon_0$ denotes the excess of mass in the centre of the galaxy.

For the purpose of Schwarzschild modelling we created libraries containing 1200 orbits (100 values of energy in units
of the radius of the circular orbit sampled logarithmically and 12 values of the relative angular momentum $l=L/L_{\rm
max}$, where $L_{\rm max}$ is the angular momentum of the circular orbit, sampled linearly) with apocentres in the
range [0.04\,:\,3.34]\,kpc. Orbits were integrated in the gravitational potential generated by the total mass
distribution dependent on $a$ and $\Upsilon_0$. We used $a\in$\,[0\,:\,2.4] with a step of $\Delta a$=0.05 and
$\Upsilon_0=1$ whereas other values of $\Upsilon_0$ were obtained with a simple algebraic transformation (see
\citealt{kowalczyk_2018}).

We fitted each orbit library to the data by minimizing the objective function $\chi^2$ given as:
 \begin{equation}
\label{eq:fit}
 \chi^2=\sum_{l}\sum_{n}\Bigg(\frac{M_l^{\rm obs}m_{n,l}^{\rm obs}-\sum_k\gamma_kM_l^km_{n,l}^{k}}{\Delta
(M_l^{\rm obs}m_{n,l}^{\rm obs})}\Bigg)^2
\end{equation}
with weights $\gamma_k$ imposed on orbits and under the constraints that for each orbit $k$ and each bin $l$:
\begin{equation}
\label{eq:weights}
\left\{
\begin{array}{l}
|M_l^{\rm obs}-\sum_k\gamma_kM_l^k|\leq\Delta M_l^{\rm obs}\\
\gamma_k\ge 0\\
\sum_k\gamma_k=1
\end{array} \right.
\end{equation}
where $M_l^k$, $M_l^{\rm obs}$ are the fractions of the projected mass of the tracer contained within the $l$th bin for
the $k$th orbit or from the observations and $m_{n,l}^k$, $m_{n,l}^{\rm obs}$ are $n$th proper moments. $\Delta$ denotes
the measurement uncertainty associated with a given parameter. The velocity moments were weighted with the projected
masses and to derive the errors we treated both quantities as independent.
We executed the $\chi^2$ fitting with rigid constraints with the
non-negative quadratic programming (QP) implemented in the CGAL library \citep{cgal}.

\section{Results}
\label{results}

\begin{figure}
\begin{center}
\includegraphics[width=\columnwidth, trim=0 0 0 0, clip]{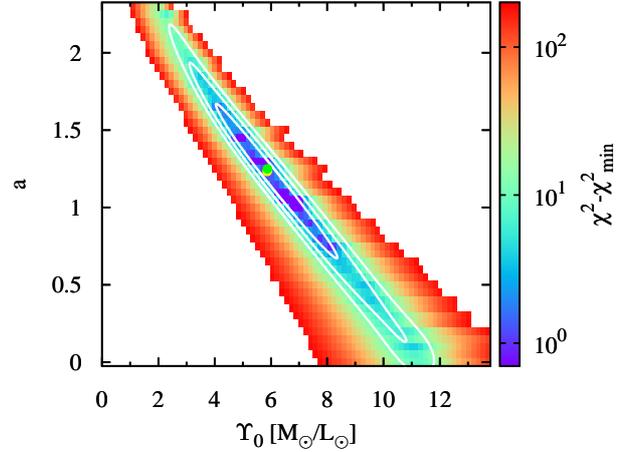}
\caption[$\Delta\chi^2$ map of Schwarzschild models for the Fornax dSph]{Colour map of absolute values of the objective
function $\chi^2$ (relative to the global minimum derived from the fitted surface) on the grid of different
mass-to-light ratio models. The minimum is marked with the yellow dot whereas the green dot indicates the best-fitting
mass profile, i.e. the profile on the grid closest to the global minimum along the contours of equal $\Delta \chi^2$
plotted with white curves.}
\label{fig:chi_fornax}
\end{center}
\end{figure}

As a result of our Schwarzschild modelling we obtained a map of absolute values of the objective function $\chi^2$ as a
function of the two parameters of the mass-to-light ratio profile: $a$ and $\Upsilon_0$. Similarly to the procedure
undertaken in our previous works, we fitted a two-dimensional 8th order ($\sim a^4\Upsilon_0^4$) surface to
the map. We derived the minimum of the surface and present the resulting $\Delta\chi^2=\chi^2-\chi^2_{\rm min}$ in
Figure~\ref{fig:chi_fornax} with the colour scale. We marked the minimum with a yellow dot and the 1,\,2,\,3$\sigma$
levels with white curves. As the numerical minimum does not correspond to any model on the adopted grid, we identified
the best-fitting mass model as a set of parameters closest to the minimum along the confidence level. It is shown in
Figure~\ref{fig:chi_fornax} as a green dot.

The $\chi^2$ map shows that the data favour high values of the curvature parameter $a$, indicating that the excess of
mass in Fornax is significantly higher at large radii. It can be explained with the presence of an extended dark matter
halo. It is worth pointing out that the mass-follows-light model, i.e. the model assuming that the spatial distribution
of the total mass is the same as the distribution of light, which for our parametrization corresponds to $a=0$, is
consistent with the data at the 3$\sigma$ confidence level. However, this model requires a high value of the
mass-to-light ratio, $\Upsilon_0\in$\,[10.6\,:\,11.7]. Since it is significantly higher than stellar mass-to-light
ratios estimated for dwarf spheroidals \citep{mateo_1998}, such a mass-follows-light model also supports the conclusion
that Fornax is embedded in a heavy dark matter halo.

\begin{figure}
\begin{center}
\includegraphics[width=\columnwidth, trim=0 0 0 0, clip]{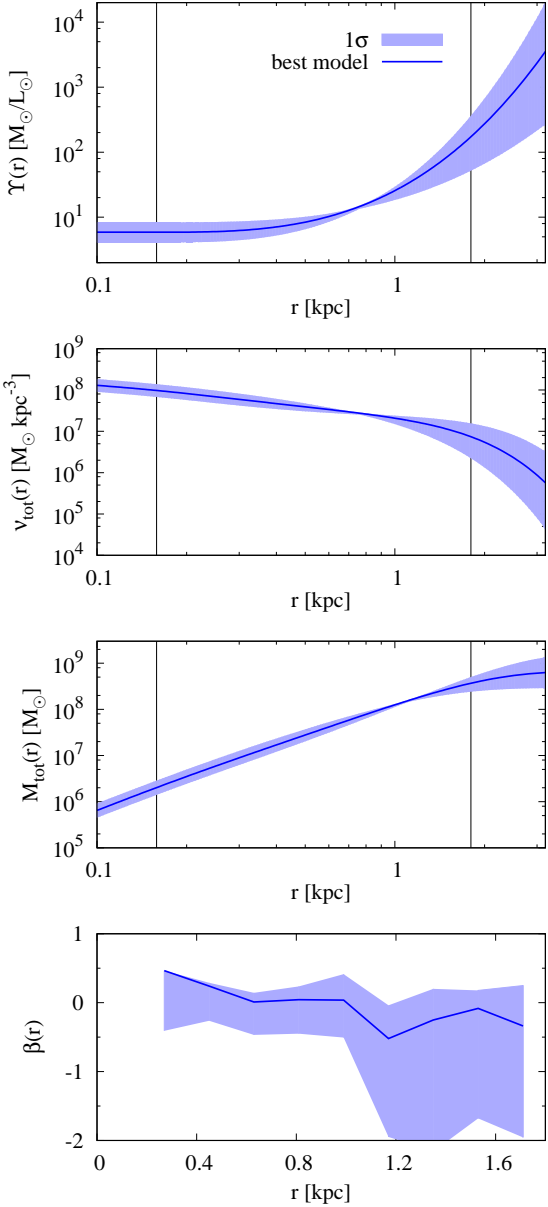}
\caption[Results of the Schwarzschild modelling of the Fornax dSph]{Results of Schwarzschild modelling of the Fornax
dSph based on the fitted mass-to-light ratio profiles.
\textit{In consecutive panels:} profiles of the mass-to-light ratio, total density, cumulative total mass and
anisotropy. Blue solid lines show the results for the best-fitting mass profile (green dot in Figure~
\ref{fig:chi_fornax}) whereas shaded regions denote the spread of values for the models from within 1$\sigma$
confidence level. Black vertical lines mark the inner radius of the luminosity profile and the outer radius of the data
used in modelling (from left to right, respectively).}
\label{fig:Upsilon}
\end{center}
\end{figure}

The derived best-fitting model and the 1$\sigma$ confidence region allowed us to construct
Figure~\ref{fig:Upsilon} where in consecutive panels we present the profiles of the mass-to-light ratio, the
total mass density, the cumulative total mass and the velocity anisotropy. Results for the best-fitting model are
presented with a blue solid line whereas the shaded areas denote the spread of values among the density models
contained within the 1$\sigma$ region (the innermost elongated ellipse in Figure~\ref{fig:chi_fornax}). Black vertical
lines mark the inner radius of the luminosity profile and the outer radius of kinematic data from left to right,
respectively.

We note that the mass profile is derived with remarkably small uncertainties, even at the outskirts of the galaxy and
in particular the mass enclosed within the radius of 1\,kpc is determined very precisely. This scale is not accidental
as \citet{wolf_2010} identified the existence of a radius (dependent on the light profile of an object) at which the
enclosed mass is almost independent of the velocity anisotropy. A rough estimate of this radius for the Fornax
luminosity distribution gives a value close to 1\,kpc (see appendix B in \citealt{wolf_2010}).

The anisotropy profile is constrained with much lower accuracy, especially at larger distances where the 1$\sigma$
confidence region is rather wide. The profile derived for the best-fitting mass model is decreasing with radius,
slightly radial ($\beta > 0$) in the centre and tangential ($\beta < 0$) at larger radii. However, the mean value is
close to isotropic, $\bar{\beta}=-0.04$.

\begin{figure}
\begin{center}
\includegraphics[width=\columnwidth, trim=0 0 0 0, clip]{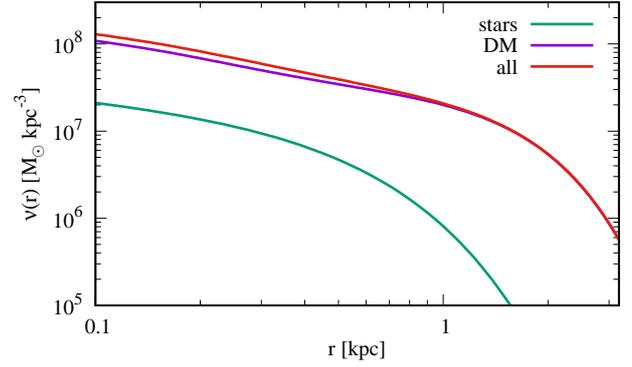}
\caption{A comparison of density profiles of stars (in green), dark matter (in purple) and total mass (in red) for
the best-fitting mass-to-light ratio model.}
\label{fig:decomp}
\end{center}
\end{figure}

In \citet{kowalczyk_2018} we demonstrated the existence of a systematic bias in the results of our spherically symmetric
modelling caused by the elliptical shape of the studied object.
We showed that for the observations
along the longest axis the derived anisotropy profile was growing, however the values of anisotropy were systematically
underestimated with the mean offset of $\sim 0.4$ for the best-fitting model, whereas for the observations
along the shortest axis the best-fitting model was consistent with isotropic orbits but the wide 1\,$\sigma$
confidence level allowed the anisotropy profile to grow (with the mean minimal offset with respect to the true values
$\Delta \beta\approx 0.1$) or decrease with radius. Since the simulated galaxy we used in that work as well
as mock data obtained with observations along the shortest axis of the stellar component are similar to Fornax
(assuming its prolate shape), we should expect an analogous bias in the present results.
Comparing the bottom panels of Figure~\ref{fig:Upsilon} and figure~8 in \citet{kowalczyk_2018} (where the blue colour
denotes the results of interest) we can see that the anisotropy inferred for Fornax is consistent with the real
anisotropy profile growing with radius from 0 to 0.5 but biased as a result of spherically symmetric modelling of a
spheroidal object observed along the minor axis.

For the sake of completeness and to better illustrate the derived model, in Fig.~\ref{fig:decomp} we also present a
comparison of density profiles of stars (in green), dark matter (in purple) and total mass (in red) for the
best-fitting mass-to-light ratio model.

\section{Summary and discussion}
\label{discussion}

Using the observational data of \citet{delpino_2015} and \citet{delpino_2017} and our version of the Schwarzschild
orbit superposition method \citep{kowalczyk_2017, kowalczyk_2018} we constructed a full dynamical model of the
Fornax dSph. We parametrized the mass content with the mass-to-light ratio varying with radius
from the centre of the galaxy described by the two parameters formula given by Eq.\,(\ref{eq:Upsilon}). We obtained
the total mass profile and inferred the presence of an extended dark matter halo. We estimated the mass contained within
the outer boundary of our kinematic data set to be $M_{\rm tot}(1.8\,$kpc$)=3.7^{+1.4}_{-1.3} \times 10^8\,$M$_{\sun}$.

Additionally, our Schwarzschild approach allowed us to derive the unparametrized velocity anisotropy profile. We
obtained nearly isotropic orbits in the centre of the galaxy and mildly tangential ones at the outskirts, however the
uncertainties grew strongly with radius.

Despite the complicated chemodynamical structure of Fornax
\citep{walker_2011, amorisco_2012, delpino_2015, delpino_2017} in this work we applied only one stellar population.
While it is our intention to implement multiple populations in our Schwarzschild code in the future, careful tests on
mock data are needed in order to maintain satisfying quality of results. Therefore, it is beyond the scope of the present
paper. Nevertheless, when deriving the light distribution we limited our photometric sample to stars of the red giant
branch, horizontal branch and red clump, to be consistent with the population from which the kinematic data originate.

\begin{figure*}
\begin{center}
\includegraphics[width=1.6\columnwidth]{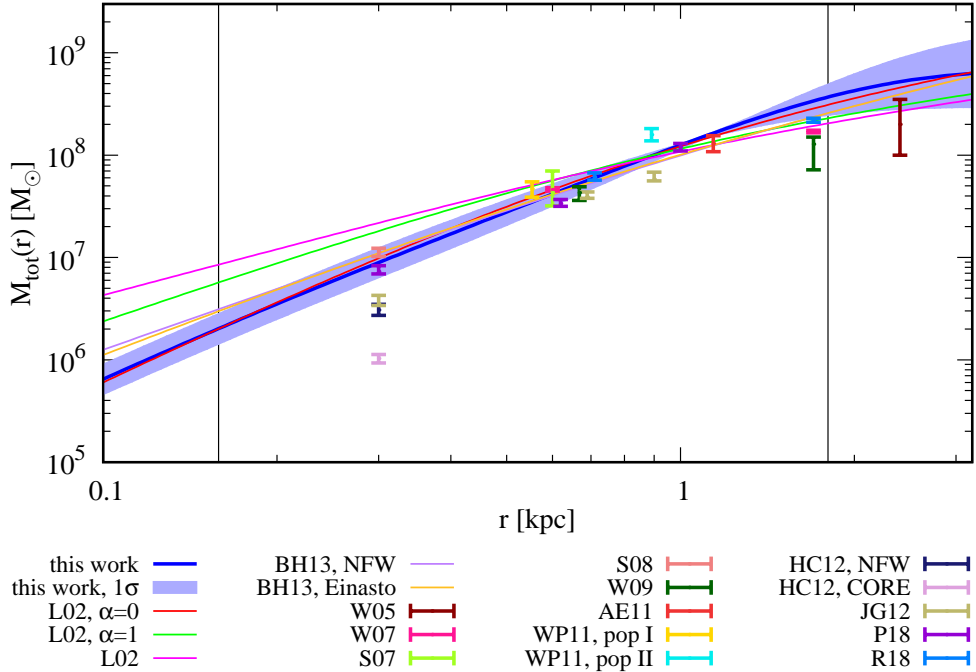}
\caption[Comparison of mass profiles and estimates for the Fornax dSph]{The comparison of derived mass profile (blue
line and light blue shaded 1$\sigma$ confidence level) with the profiles (other colour lines) and estimates (error
bars) from the literature. For the estimates we present only 1$\sigma$ error bars (without data points) to avoid
crowding.\\
\textit{Acronyms used:} L02 \citet{lokas_2002}, BH13 \citet{breddels_2013b}, W05 \citet{wang_2005}, W07
\citet{walker_2007}, S07 \citet{strigari_2007}, S08 \citet{strigari_2008}, W09 \citet{walker_2009b}, AE11
\citet{amorisco_2011}, WP11 \citet{walker_2011}, HC12 \citet{hayashi_2012}, JG12 \citet{jardel_2012}, P18
\citet{pascale_2018}, R18 \citet{read_2018}}
\label{fig:comp}
\end{center}
\end{figure*}

In Figure~\ref{fig:comp} we present an extensive comparison of our estimated mass profile for Fornax with
different results from the literature. The blue line and the light blue shaded area represent the profile and the
1$\sigma$ confidence region obtained in this work. For the results from the literature we applied the following
approach: 1) if the parameters of the models were given or could be read from the figures, we reproduce full mass
profiles, 2) if a full profile was derived but is difficult to reproduce, the values at specific radii given by the
authors are presented (for \citealt{read_2018} the values of the dark matter mass were read from a
figure and we used our values of the stellar mass at given radii to calculate the total masses), 3) values based on
estimators are shown at appropriate radii. In order to avoid crowding, in the first case we draw just the profiles with
the best-fitting parameters whereas for the other cases we present only 1$\sigma$ error bars (without data points). The
references are given in the caption of Figure~\ref{fig:comp}.

{\renewcommand{\arraystretch}{1.3}
\begin{table*}
\begin{minipage}{120mm}
\caption{Comparison of the derived values of anisotropy (both constant and varying with radius).}
\begin{threeparttable}[b]
\label{tab:fornax_anisotropy}\ \\
\begin{tabular}{lcccc}
\hline
\multicolumn{5}{c}{$\beta=$\,const}\\
\hline
& & & model & $\beta$\\
\hline
\multirow{3}{*}{\citet{lokas_2002}} & & & $\alpha=0$ & $-0.36\pm$0.33\\
& & & $\alpha=1$ & $-1.5\pm$0.8\\
& & & $\alpha=1.5$ & $-2.6\pm$1.6\\
\citet{walker_2007} & & & & $-0.5$\\
\citet{walker_2009b} & & & & $-0.6$\\
\hline
\hline
\multicolumn{5}{c}{$\beta (r)$}\\
\hline
& & shape & $\beta(0.3$\,kpc) & $\beta (1.7$\,kpc)\\
\hline
this work & & decreasing & 0.46$^{+0.00}_{-0.87}$ & $-0.34^{+0.59}_{-1.62}$ \\
\citet{hayashi_2012} & & local max \tnote{1} & 0.09 & 0.26 \\
\multirow{2}{*}{\citet{jardel_2012}} & major axis & growing & 0.17$^{+0.19}_{-0.55}$ & 0.49$^{+0.16}_{-0.25}$ \tnote{2} \\
& minor axis & growing & $-0.38^{+0.27}_{-0.40}$ & 0.52$^{+0.09}_{-0.21}$ \tnote{2} \\
\citet{breddels_2013b} & & flat & \multicolumn{2}{c}{0.2$\pm$0.2} \\
\citet{pascale_2018} & & flat & 0.05$^{+0.05}_{-0.05}$ & 0.1$^{+0.15}_{-0.20}$ \\
\citet{read_2018} & & flat & $-0.26^{+0.16}_{-0.14}$ & $-0.18^{+0.08}_{-0.15}$ \tnote{3} \\
\hline
\end{tabular}
\begin{tablenotes}
 \item[1] $\beta (1$\,kpc)=0.32
 \item[2] Values at the outermost point $r\approx 0.92$\,kpc.
 \item[3] Values at the outermost point $r=1$\,kpc.
\end{tablenotes}
\end{threeparttable}
\end{minipage}
\end{table*}} \quad

When considering the full mass profile, our result agrees very well at all radii with the profile derived by
\citet{lokas_2002} for the cored model $\alpha=0$, where $\alpha$ denotes the asymptotic central slope for a
generalized Navarro-Frenk-White profile (NFW, \citealt{NFW_1997}) and generally disagrees with models for $\alpha=1,
1.5$. This is not surprising since in our modelling we assumed that in the centre the dark matter slope follows that of
the stellar component and they have the central slope of 0.32. Cuspy profiles derived by \citet{breddels_2013b}: NFW
and Einasto \citep{einasto_1965} give very similar mass profiles which lie within our 1$\sigma$ confidence level,
but fall a little below our result at radii where our profile has the narrowest confidence region.

{\renewcommand{\arraystretch}{1.3}
\begin{table*}
 \begin{minipage}{112mm}
\caption{Comparison of total masses and values of the anisotropy for mass-follows-light models.}
\begin{threeparttable}[b]
\label{tab:fornax_ml}\ \\
\begin{tabular}{lccc}
\hline
& $M_{\infty}$\,[$\times 10^8$\,M$_{\sun}$] & $\bar{\beta}$ & $\beta=$\,const\\
\hline
this work & 1.74$\pm$0.04 & $-1.94^{+0.11}_{-0.03}$ \tnote{1} & \\
\multirow{2}{*}{\citet{klimentowski_2007} \tnote{2}}
& 2.03$^{+0.38}_{-0.32}$ & & $-0.17^{+0.37}_{-0.63}$ \\
& 1.91$^{+0.32}_{-0.31}$ & & $-1.82^{+1.02}_{-2.66}$ \\
\citet{lokas_2009} & 1.57$\pm$0.07 & & $-0.32^{+0.14}_{-0.17}$ \\
\citet{diakogiannis_2017}& $1.613^{+0.050}_{-0.075}$ & $-0.95^{+0.78}_{-0.72}$ & \\
\hline
\end{tabular}
\begin{tablenotes}
 \item[1] Mean weighted with deprojected light profile.
 \item[2] Rows correspond to different sample cleaning procedures.
\end{tablenotes}
\end{threeparttable}
 \end{minipage}
\end{table*}} \quad

Our mass profile is consistent within 1$\sigma$ confidence level with most of the results presented with the error
bars. We can distinguish three types of discrepancies: at large radii ($r > 1.5$\,kpc) as they correspond to outer
boundaries of data sets (this work, \citealt{wang_2005, walker_2007, walker_2009b}), with the result for the metal-poor
population (pop II) from \citet{walker_2011} and with results of axisymmetric models by \citet{hayashi_2012} and
\citet{jardel_2012}. Whereas the low values obtained by \citet{hayashi_2012} can be caused by the assumption of the
non-spherical shape of the dark matter halo, the interpretation of systematically lower masses given by
\citet{jardel_2012} is more complicated. In \citet{kowalczyk_2018} we showed that the results for observations of a
galaxy along the shortest axis are not biased whereas those for the longest axis are overestimated. Assuming that the
orientation of Fornax is between these extreme cases, our profile should be slightly overestimated. On the other hand,
values obtained with mass estimators are on average unbiased \citep{kowalczyk_2013, campbell_2017} and since they are
in good agreement with our profile at various radii, it suggests that bias on our results is rather insignificant. We
will return to the results of axisymmetric modelling when comparing the derived anisotropy profiles. Interestingly, our
model is in good agreement with the results of \citet{pascale_2018} who partially took into account the ellipticity by
assigning to each star a `circularized radius' dependent on both coordinates and the flattening.

In Table~\ref{tab:fornax_anisotropy} we compare the values of the derived anisotropy. The top part of the Table refers
to the models with the anisotropy assumed to be constant with radius whereas in the bottom part we present the values
for models varying with radius. We show the values at two radii: 0.3\,kpc and 1.7\,kpc which correspond to the inner and
outer radii of our anisotropy profile (see Figure~\ref{fig:Upsilon}). We also indicate the general shape of the
profile.

The values of the anisotropy from \citet{walker_2007} and the profiles varying with radius (except for
\citealt{breddels_2013b}) were read from the figures published in these papers and are therefore approximate.
We note that the profiles of mass distribution and
anisotropy in \citet{jardel_2012} are shown as a function of radius given in arcseconds so the outer radius of the
anisotropy profile should correspond to about 0.92\,kpc (as given in Table~\ref{tab:fornax_anisotropy}).
However, after careful analysis of the text and given our experience with Fornax, we have reasons to believe
that the profiles in \citet{jardel_2012} are really plotted as a function of the physical radius (in kpc) with the outer
boundary of $\sim 1.6$\,kpc.

The values of anisotropy assumed to be constant with radius (top part of
Table~\ref{tab:fornax_anisotropy}) from the literature are consistently negative whereas our mean anisotropy is close to
zero. This discrepancy, except for the result in \citet{lokas_2002} for $\alpha=0$, can be explained with the
mass-anisotropy degeneracy as the corresponding mass estimates are lower at large radii (see Figure~\ref{fig:comp}).
Moreover, despite of deriving full anisotropy profiles, \citet{breddels_2013b} and \citet{pascale_2018} also obtained
flat models but with higher values, closer to ours.

As already stated in Section~\ref{results}, our derived anisotropy profile is consistent with a growing profile biased
by the ellipticity of the galaxy and its observation along the shortest axis. Interestingly, such a growing profile
would agree with the findings of \citet{jardel_2012} (see Table~\ref{tab:fornax_anisotropy}) who, contrary to other
authors, obtained anisotropy increasing with radius. Instead of spherically symmetric modelling, they applied an
axisymmetric approach but assumed a particular orientation of the galaxy, avoiding an additional parameter.
Even more general axisymmetric models were applied by \citet{hayashi_2012} and the dominant trend in their anisotropy
profile is also increasing. It may suggest that their treatment of the elliptical projected shape of the galaxy was
able to lift the bias.

Finally, in Table~\ref{tab:fornax_ml} we compare our results for the mass-follows-light model, i.e. with $a=0$, with the
total mass and anisotropy estimates for the same type of mass distribution from the literature. Since the light profiles
and the total luminosity estimates differ between the studies, we present the derived total masses. All the results are
consistent with our findings within 1$\sigma$ confidence level.

Results for the anisotropy are less conclusive. Since our method recovers the full anisotropy profile (instead of just
a constant value) we present the range of mean values of the anisotropy. They were calculated by weighting the
radius-dependent quantities with the deprojected stellar mass fractions as the values of anisotropy rapidly decrease
from $\approx 0$ at the centre to $\approx -7$ at the outer boundary of the data set ($r=1.8$\,kpc). The derived
profiles are approximately consistent with the results presented by \citet{diakogiannis_2017} who also derive the full
anisotropy profile. Similarly to our findings, they obtained the profile decreasing from 0 in the centre to $\approx
-1.5$ at 1\,kpc. However, their profile has a minimum of $-2$ at 1.5\,kpc (which was the outer radius of the data set)
and grows beyond. The mean value cited in the Table is as given by \citet{diakogiannis_2017}. For the models with
constant anisotropy, our result agrees well with the value derived by \citet{klimentowski_2007} for a less restrictive
procedure of data sample cleaning but strongly disagrees with their other value and findings of \citet{lokas_2009}. We
note that the latter two studies used a much more restrictive sample cleaning method than our conservative $3\sigma$
clipping. Nevertheless, all studies reproduce a tangentially biased anisotropy. Since our modelling suggests the
existence of an extended dark matter halo, mass-follows-light models underestimate the mass content at large radii and
necessarily yield lower, more tangential, values of anisotropy in order to recreate the same velocity distribution with
the Jeans equation.

\section*{Acknowledgements}

This research was supported in part by the
Polish Ministry of Science and Higher Education under grant 0149/DIA/2013/42 within the Diamond Grant Programme for
years 2013-2017 and by the Polish National Science Centre under grant 2013/10/A/ST9/00023. AdP acknowledges support of
and discussions with the HSTPROMO collaboration. MV acknowledges support from HST-AR-13890.001, NSF awards AST-0908346,
AST-1515001, NASA-ATP award NNX15AK79G.

\bsp
\label{lastpage}

\begin{thebibliography}{}
\bibitem[\protect\citeauthoryear{{Amorisco} \& {Evans}}{{Amorisco} \&
{Evans}}{2011}]{amorisco_2011}
{Amorisco} N.~C., {Evans} N.~W., 2011, \mnras, 411, 2118

\bibitem[\protect\citeauthoryear{Amorisco \& Evans}{2012}]{amorisco_2012}
Amorisco N.~C., Evans N.~W., 2012, ApJ, 756, L2

\bibitem[\protect\citeauthoryear{Amorisco, Agnello \& Evans}{Amorisco et~al.}{2013}]{amorisco_2013}
Amorisco N.~C., Agnello A., Evans N.~W., 2013, MNRAS, 429, L89

\bibitem[\protect\citeauthoryear{{Battaglia}, {Sollima}, \& {Nipoti}}{{Battaglia}
et~al.}{2015}]{battaglia_2015}
{Battaglia} G., {Sollima} A., {Nipoti} C., 2015, \mnras, 454, 2401

\bibitem[\protect\citeauthoryear{Battaglia et al.}{Battaglia
 et al.}{2006}]{battaglia_2006}
Battaglia G. et~al., 2006, \aap, 459, 423

\bibitem[\protect\citeauthoryear{Binney \& Mamon}{Binney \& Mamon}{1982}]{binney_1982}
Binney J., Mamon G.~A., 1982, \mnras, 200, 361

\bibitem[\protect\citeauthoryear{Binney \& Tremaine}{Binney \&
  Tremaine}{2008}]{GD}
Binney J.,  Tremaine S.,  2008, Galactic Dynamics, 2 edn.
Princeton University Press, Princeton, NJ

\bibitem[\protect\citeauthoryear{{Bouchard}, {Carignan}, \& {Staveley-Smith}}
{{Bouchard} et~al.}{2006}]{bouchard_2006}
{Bouchard} A., {Carignan} C., {Staveley-Smith} L., 2006, \aj, 131, 2913

\bibitem[\protect\citeauthoryear{{Campbell} et~al.}{{Campbell} et~al.}
{2017}]{campbell_2017}
{Campbell} D.~J.~R. et~al., 2017, \mnras, 469, 2335

\bibitem[\protect\citeauthoryear{\lowercase{De} Boer et al.}
{\lowercase{De} Boer et~al.}{2012}]{deboer_2012}
{\lowercase{De} Boer T.~J.~L.} et~al., 2012, \aap, 544, A73

\bibitem[\protect\citeauthoryear{Breddels \& Helmi}{Breddels \&
Helmi}{2013}]{breddels_2013b}
Breddels M.~A., Helmi A., 2013, \aap, 558, A35

\bibitem[\protect\citeauthoryear{\lowercase{Del} Pino et al.}
{\lowercase{Del} Pino et~al.}{2013}]{delpino_2013}
{del Pino} A., {Hidalgo} S.~L., {Aparicio} A., {Gallart} C.,
{Carrera} R., {Monelli} M., {Buonanno} R., {Marconi} G., 2013,
\mnras, 433, 1505

\bibitem[\protect\citeauthoryear{\lowercase{Del} Pino, Aparicio \& Hidalgo}
{\lowercase{Del} Pino et~al.}{2015}]{delpino_2015}
\lowercase{Del} Pino A., Aparicio A., Hidalgo S.~L.,  2015, \mnras, 454, 3996

\bibitem[\protect\citeauthoryear{\lowercase{Del} Pino et al.}
{\lowercase{Del} Pino et~al.}{2017}]{delpino_2017}
\lowercase{Del} Pino A., Aparicio A., Hidalgo S.~L., {\L}okas E.~L., 2017,
\mnras, 465, 3708

\bibitem[\protect\citeauthoryear{Diakogiannis et~al.}
{Diakogiannis et~al.}{2017}]{diakogiannis_2017}
Diakogiannis F.~I. et~al., 2017, \mnras, 470, 2034

\bibitem[\protect\citeauthoryear{Einasto}{Einasto}{1965}]{einasto_1965}
Einasto J., 1965, Trudy Astrofizicheskogo Instituta Alma-Ata, 5, 87

\bibitem[\protect\citeauthoryear{{Hayashi} \& {Chiba}}{{Hayashi} \&
{Chiba}}{2012}]{hayashi_2012}
{Hayashi} K., {Chiba} M., 2012, \apj, 755, 145

\bibitem[\protect\citeauthoryear{Hayashi \& Chiba}{2015}]{hayashi_2015}
Hayashi K., Chiba M., 2015, ApJ, 810, 22

\bibitem[\protect\citeauthoryear{{Irwin} \& {Hatzidimitriou}}{{Irwin} \&
{Hatzidimitriou}}{1995}]{irwin_1995}
{Irwin} M., {Hatzidimitriou} D., 1995, \mnras, 277, 1354

\bibitem[\protect\citeauthoryear{Jardel \& Gebhardt}{Jardel \&
Gebhardt}{2012}]{jardel_2012}
Jardel J.~R., Gebhardt K., 2012, \apj, 746, 89

\bibitem[\protect\citeauthoryear{Jardel \& Gebhardt}{2013}]{jardel_2013}
Jardel J.~R., Gebhardt K., 2013, ApJ, 775, L30

\bibitem[\protect\citeauthoryear{Jardel, Gebhardt, Fabricius, Drory \& Williams}{Jardel et~al.}
{2013}]{jardel_2013a}
Jardel J.~R., Gebhardt K., Fabricius M.~H., Drory N., Williams M.~J., 2013, ApJ, 763, 91

\bibitem[\protect\citeauthoryear{Kirby et~al.}{Kirby
  et~al.}{2010}]{kirby_2010}
Kirby E.~N. et~al.,  2010, \apjs, 191, 352

\bibitem[\protect\citeauthoryear{{Klimentowski} et~al.}{{Klimentowski} et~al.}{2007}]{klimentowski_2007}
{Klimentowski} J., {{\L}okas} E.~L., {Kazantzidis} S., {Prada} F.,
{Mayer} L., {Mamon} G.~A., 2007, \mnras, 378, 353

\bibitem[\protect\citeauthoryear{Lima, Gerbal \& M\'arquez}
{Lima, Gerbal \& M\'arquez}{1999}]{lima_1999}
Lima Neto G.~B., Gerbal D., M\'arquez I.,  1999, \mnras, 309, 481

\bibitem[\protect\citeauthoryear{Kendall \& Stuart}{Kendall \&
  Stuart}{1977}]{kendall_1977}
Kendall M.,  Stuart A.,  1977, The advanced theory of statistics, Vol 1, 4 edn.
Charles Griffin \& Co Ltd, London \& High Wycombe, UK

\bibitem[\protect\citeauthoryear{{Kowalczyk}, {{\L}okas}, {Kazantzidis}, \&
  {Mayer}}{{Kowalczyk} et~al.}{2013}]{kowalczyk_2013}
{Kowalczyk} K., {{\L}okas} E.~L., {Kazantzidis} S., {Mayer} L.,
2013, \mnras, 431, 2796

\bibitem[\protect\citeauthoryear{Kowalczyk, {\L}okas  \& Valluri}{Kowalczyk
  et~al.}{2017}]{kowalczyk_2017}
Kowalczyk K.,  {\L}okas E.~L.,   Valluri M.,  2017, \mnras, 470, 3959

\bibitem[\protect\citeauthoryear{Kowalczyk, {\L}okas  \& Valluri}{Kowalczyk
  et~al.}{2018}]{kowalczyk_2018}
Kowalczyk K.,  {\L}okas E.~L.,   Valluri M.,  2018, \mnras, 476, 2918

\bibitem[\protect\citeauthoryear{{Letarte} et~al.}{{Letarte} et~al.}
{2010}]{letarte_2010} {Letarte} B. et~al., 2010, \aap, 523, A17

\bibitem[\protect\citeauthoryear{{\L}okas}{{\L}okas}{2002}]{lokas_2002}
{\L}okas E.~L., 2002, \mnras, 333, 697

\bibitem[\protect\citeauthoryear{{\L}okas}{{\L}okas}{2009}]{lokas_2009}
{\L}okas E.~L., 2009, \mnras, 394, 102

\bibitem[\protect\citeauthoryear{{\L}okas \& Mamon}{{\L}okas \&
  Mamon}{2003}]{lokas_2003}
{\L}okas E.~L.,  Mamon G.~A.,  2003, \mnras, 343, 401

\bibitem[\protect\citeauthoryear{Mateo}{Mateo}{1998}]{mateo_1998}
Mateo M.,  1998, \araa, 36, 435

\bibitem[\protect\citeauthoryear{McConnachie}{McConnachie}{2012}]{mcconnachie_2012}
{McConnachie} A.~W., 2012, \aj, 144, 4

\bibitem[\protect\citeauthoryear{Navarro, Frenk  \& White}{Navarro
  et~al.}{1997}]{NFW_1997}
Navarro J.~F.,  Frenk C.~S.,   White S. D.~M.,  1997, \apj, 490, 493

\bibitem[\protect\citeauthoryear{{Pascale}, {Posti}, {Nipoti} \&
{Binney}}{{Pascale} et~al.}{2018}]{pascale_2018}
{Pascale} R., {Posti} L., {Nipoti} C., {Binney} J., 2018, MNRAS, 480, 927

\bibitem[\protect\citeauthoryear{Read, Walker \& Steger}{Read et~al.}{2018}]{read_2018}
Read J.~I., Walker M.~G., Steger P., 2018, ArXiv e-prints, arXiv:1808.06634

\bibitem[\protect\citeauthoryear{Sanders, Evans \& Dehnen}{Sanders et~al.}
{2018}]{sanders_2018}
Sanders J.~L., Evans N.~W., Dehnen W., 2018, MNRAS, 478, 3879

\bibitem[\protect\citeauthoryear{Sanders, Evans, Geringer-Sameth \& Dehnen}{Sanders et~al.}
{2016}]{sanders_2016}
Sanders J.~L., Evans N.~W., Geringer-Sameth A., Dehnen W., 2016, PhRvD, 94, 63521

\bibitem[\protect\citeauthoryear{Schwarzschild}{Schwarzschild}{1979}]{schwarzschild_1979}
Schwarzschild M.,  1979, \apj, 232, 236

\bibitem[\protect\citeauthoryear{S\'ersic}{S\'ersic}{1968}]{sersic_1968}
S\'ersic J.~L., 1968, Atlas de galaxias australes, Observatorio Astronomico,
Cordoba

\bibitem[\protect\citeauthoryear{Stetson}{Stetson}{2000}]{stetson_2000}
Stetson P.~B.,  2000, \pasp, 112, 925

\bibitem[\protect\citeauthoryear{Stetson}{Stetson}{2005}]{stetson_2005}
Stetson P.~B.,  2005, \pasp, 117, 563

\bibitem[\protect\citeauthoryear{{Strigari} et~al.}
{{Strigari} et~al.}{2007}]{strigari_2007}
{Strigari} L.~E., {Bullock} J.~S., {Kaplinghat} M., {Diemand} J.,
{Kuhlen} M., {Madau} P., 2007, \apj, 669, 676

\bibitem[\protect\citeauthoryear{{Strigari} et~al.}
{{Strigari} et~al.}{2008}]{strigari_2008}
{Strigari} L.~E., {Bullock} J.~S., {Kaplinghat} M., {Simon} J.~D., {Geha} M.,
{Willman} B., {Walker} M.~G., 2008, \nat, 454, 1096

\bibitem[\protect\citeauthoryear{{The CGAL Project}}{{The CGAL
  Project}}{2015}]{cgal}
{The CGAL Project} 2015, {CGAL} User and Reference Manual, {4.7} edn.
{CGAL Editorial Board}, \url {http://doc.cgal.org/4.7/Manual/packages.html}

\bibitem[\protect\citeauthoryear{Walker~et~al.}{Walker
et~al.}{2007}]{walker_2007}
Walker M.~G., Mateo M., Olszewski E.~W., Gnedin O.~Y., Wang X.,
Sen B., Woodroofe M., 2007, \apjl, 667, 56

\bibitem[\protect\citeauthoryear{Walker, Mateo \& Olszewski}{Walker
et~al.}{2009a}]{walker_2009a}
Walker M.~G.,  Mateo M.,  Olszewski E.~W.,  2009a, \aj, 137, 3100

\bibitem[\protect\citeauthoryear{Walker et~al.}{Walker et~al.}{2009b}]{walker_2009b}
Walker M.~G., et~al., 2009b, \apj, 704, 1274

\bibitem[\protect\citeauthoryear{{Walker} \& {Pe{\~n}arrubia}}{{Walker} \&
{Pe{\~n}arrubia}}{2011}]{walker_2011}
{Walker} M.~G., {Pe{\~n}arrubia} J., 2011, \apj, 742, 20

\bibitem[\protect\citeauthoryear{{Wang} et~al.}{{Wang} et~al.}{2005}]{wang_2005}
{Wang} X., {Woodroofe} M., Walker M.~G., Mateo M., Olszewski E.,
2005, \apj, 626, 145

\bibitem[\protect\citeauthoryear{Wolf et~al.}{Wolf et~al.}{2010}]{wolf_2010}
Wolf, J., et~al., 2010, \mnras, 406, 1220
\end{thebibliography}
\end{document}